\documentclass[11pt,prd,onecolumn,showpacs,amsmath,amssymb,aps,floats,floatfix]{revtex4-1}
\usepackage{multirow}
\usepackage{graphics,graphicx,subfigure}
\usepackage{pstricks}
\graphicspath{{figs/}}  
\usepackage{mathrsfs,amsmath}
\usepackage{float}
\usepackage{ulem}

\def\sigmav{\langle \sigma v \rangle}
\def\sigmav{\langle \sigma_{\mathrm{ann}} v \rangle}
\def\TeV{\mathrm{TeV}} 
\def\GeV{\mathrm{GeV}} 

\begin{document}
\title{A dark matter model that reconciles tensions between the cosmic-ray $e^\pm$ excess and the gamma-ray and CMB constraints}
\author{Qian-Fei Xiang$^{1,2}$}
\author{Xiao-Jun Bi$^1$}
\author{Su-Jie Lin$^1$}
\author{Peng-Fei Yin$^1$}
\affiliation{$^1$Key Laboratory of Particle Astrophysics,
Institute of High Energy Physics, Chinese Academy of Sciences,
Beijing 100049, China}
\affiliation{$^2$School of Physical Sciences,
University of Chinese Academy of Sciences,
Beijing 100049, China}

\begin{abstract}

The cosmic-ray (CR) $e^\pm$ excess observed by AMS-02 can be explained by dark matter (DM) annihilation. However, the DM explanation requires a large annihilation cross section which is strongly disfavored by other observations, such as the Fermi-LAT gamma-ray observation of dwarf galaxies and the Planck observation of the cosmic microwave background (CMB). Moreover, the DM annihilation cross section required by the CR $e^\pm$ excess is also too large to generate the correct DM relic density with thermal production. In this work we use the Breit-Wigner mechanism with a velocity dependent DM annihilation cross section to reconcile these tensions.
If DM particles accounting for the CR $e^\pm$ excess with $v\sim \mathcal{O}(10^{-3})$ are very close to a resonance in the physical pole case, their annihilation cross section in the Galaxy reaches a maximal value. On the other hand, the annihilation cross section would be suppressed for DM particles with smaller relative velocities in dwarf galaxies and at recombination, which may affect the gamma-ray and CMB observations, respectively. We find a proper parameter region that can simultaneously explain the AMS-02 results and the thermal relic density, while satisfying the Fermi-LAT and Planck constraints.

\end{abstract}

\pacs{95.35.+d, 98.70.Sa}

\maketitle
\section{INTRODUCTION}
Astrophysics and cosmology observations reveal that the dominant matter component in the universe is dark matter (DM), but the particle nature of DM remains unknown~\cite{Komatsu:2008hk,Ade:2015xua}.
The existence of DM cannot be explained within the framework of the standard model (SM), and thus provides a hint of the physics beyond the SM. Great efforts have been devoted to DM researches, including collider detection, direct detection, and indirect detection experiments.

DM particles can be traced by cosmic ray (CR) experiments through their annihilation products from the Galaxy halo.
The Alpha Magnetic Spectrometer (AMS-02), launched in 2011, is able to measure CR spectra with an unprecedented precision~\cite{Aguilar:2013qda}. The precise results released by AMS-02 have confirmed the CR $e^\pm$ excess above $\sim 10$ GeV, which indicates the existence of exotic $e^\pm$ sources. Many astrophysical explanations have been proposed for this excess, such as primary sources like pulsars~\cite{Hooper:2008kg,Yuksel:2008rf,Profumo:2008ms}, or the CR interactions occurring around CR acceleration sources~\cite{Blasi:2009hv,Hu:2009zzb,Fujita:2009wk}.
Interestingly, this excess can also be explained by DM annihilations/decays to charged leptons~\cite{Bergstrom:2008gr,Barger:2008su,Cirelli:2008pk,Yin:2008bs,Zhang:2008tb,Bergstrom:2009fa,Lin:2014vja}.

On the other hand, DM particles would also generate high energy photons associated with charged leptons.
The related gamma-ray signatures can be significant in systems with high DM densities and low baryon densities, such as dwarf galaxies. However, the Fermi-LAT observations do not find such signatures, and set strong constraints on the DM annihilation cross section~\cite{Ackermann:2013yva,Ackermann:2015zua,Fermi-LAT:2016uux}.
Since the large annihilation cross section required by the CR $e^\pm$ excess seems not to be allowed by the Fermi-LAT constraints~\cite{Lin:2014vja}, the DM annihilation explanation is strongly disfavored.

Moreover, the electromagnetically interacting particles generated by DM annihilations at recombination could affect cosmic microwave background (CMB)~\cite{Chen:2003gz,Padmanabhan:2005es,Slatyer:2015kla,Slatyer:2015jla,Galli:2009zc}.
Precise measurements performed by WMAP~\cite{Komatsu:2008hk} and recently by Planck~\cite{Ade:2015xua} have been used to set constraints on the DM energy injections and the DM annihilation cross sections for specified final states.
Compared to the results from CR and gamm-ray observations, these constraints are more stringent, and are free of some astrophysical uncertainties, which arise from the large-scale structure formation, DM density files and so on~\cite{Slatyer:2015jla}.

Apparently, the results from the Fermi-LAT and Planck observations strongly disfavor the large DM annihilation cross sections required by  the CR $e^\pm$ excess~\cite{Lin:2014vja}. However, note that DM particles have very different relative velocities in different circumstances. For the DM particles potentially impacting on the CR $e^\pm$, dwarf galaxy gamma-ray, and CMB observations, the typical relative velocities are $v\sim 10^{-3}$, $10^{-4}$, and $\ll 10^{-6}$, respectively. Therefore, the inconsistence between the DM explanations for different experimental results can be relaxed or even avoided by a velocity dependent annihilation cross section. In fact, the velocity dependent DM annihilation models, such as the Sommerfeld~\cite{Hisano:2003ec,Hisano:2006nn,Cirelli:2007xd,ArkaniHamed:2008qn,Feng:2009hw, Feng:2010zp, Cirelli:2016rnw} and Breit-Wigner mechanisms~\cite{Feldman:2008xs, Ibe:2008ye,Guo:2009aj,Bi:2009uj,Bi:2011qm,Bai:2017fav}, have been widely used to simultaneously explain the thermal DM relic density and the CR $e^\pm$ excess. In these models, DM particles have a much larger annihilation cross section in the Galaxy with $v\sim 10^{-3}$ than that in the early Universe for explaining the relic density with $v\sim 10^{-1}$.

In this paper, we explain the AMS-02 $e^\pm$ excess in an annihilating DM scenario with the Breit-Wigner mechanism. The DM relic density and the constraints from the Fermi-LAT and Planck observations are also taken into account. In this scenario, two DM particles resonantly annihilate via the s-channel exchange of a heavy mediator. The typical form of the DM annihilation cross section is characterized by two parameters, namely $\gamma \equiv \Gamma_{Z'}/m_{Z'}$ and $\delta \equiv 1- m_{Z'}^2 / 4 m_{\chi}^2$, where $\Gamma_{Z'}$, $m_{Z'}$, and $m_{\chi}$ are the mediator decay width, the mediator mass, and the DM mass, respectively. The assumptions of $\delta >0$ and $\delta <0$ correspond to the cases with an unphysical pole and a physical pole, respectively. As shown in
Ref.~\cite{Feldman:2008xs,Ibe:2008ye,Guo:2009aj}, both these two cases can simultaneously explain the high energy positron excess observed by PAMELA and the DM relic density. In our analysis, we perform a fitting to the AMS-02 $e^\pm$ data with the DM contribution, and derive the corresponding DM annihilation cross sections for $\mu^+\mu^-$ and $\tau^+\tau^-$ final states. Then we adjust the parameters $\gamma$ and $\delta$ to obtain suitable DM annihilation cross sections with different relative velocities. We find that there exists a parameter region with $\delta<0$, simultaneously accounting for the AMS-02 $e^\pm$ excess and DM relic density, which is also allowed by the Fermi-LAT dwarf galaxy gamma-ray and the Planck CMB observations.

This paper is organized as follows.
In Sec.~\ref{sec:fit} we perform a fitting to the AMS-02 data, and derive the corresponding DM annihilation cross sections for $\mu^+\mu^-$ and $\tau^+\tau^-$ final states.
In Sec.~\ref{sec:enhance} we briefly introduce the Breit-Wigner scenario.
In Sec.~\ref{sec:results} we show how to relax the tension between DM explanations for the AMS-02, Fermi, and Planck observations, and obtain the correct DM relic density.
Sec.~\ref{sec:conclusion} is our conclusions and discussions.

\section{Fit to the AMS-02 data}
\label{sec:fit}
The complicated CR propagation process can be described by a propagation equation involving some free parameters. In order to predict the CR $e^\pm$ background, some additional parameters describing the primary and secondary CR injections are needed. In principle, these parameters are determined by available CR observations. In this work, we use the package GALPROP~\cite{Strong:1998pw,Moskalenko:1997gh} to resolve the propagation equation, and perform a Markov chain Monte Carlo fitting to the AMS-02 data in the high dimensional parameter space.

The propagation parameters are dominantly determined by a fitting to the measured secondary-to-primary ratios \cite{Lin:2014vja}, including the B/C data from ACE~\cite{2000AIPC..528.....M} and AMS-02~\cite{2013ICRC-AMS02}, and the $^{10}\mathrm{Be}/^{9}\mathrm{Be}$ data from several experiments. Two kinds
of propagation models, namely the diffusion-convection (DR) model and the diffusion-reacceleration (DR) model, are taken into account in \cite{Lin:2014vja}.
The injection spectrum of the primary electron background is assumed to be a three-piece broken power law with two breaks.
Comparing to the spectrum with only one break at the low energy, we find that the spectrum with an additional break around $60~\GeV$ can provide a better fit to the AMS-02 data.
The nucleon injection parameters are constrained by fitting the proton flux of AMS-02~\cite{2013ICRC-AMS02}.
After deriving the propagated proton spectrum, the injection of the secondary $e^\pm$ backgrounds is calculated by using the parameterized cross section presented in Ref.~\cite{Kamae:2006bf}.

For the DM signature, we assume that DM particles purely annihilate to $\mu^+\mu^-$ or $\tau^+\tau^-$. The initial $e^\pm$ spectra from DM annihilation are calculated by PPPC 4 DM ID~\cite{Cirelli:2010xx}, which includes the electroweak corrections~\cite{Ciafaloni:2010ti}. The DM density profile is taken to be the NFW profile~\cite{Navarro:1996gj} defined by $\rho(r)=\rho_s r_s/r(1+r/r_s)^{2}$, with a characteristic halo radius $\rho_s = 20~\mathrm{kpc}$ and a characteristic halo density $\rho_s = 0.26~\mathrm{GeV cm^{-3}}$.

\begin{figure}[!htbp]
	\subfigure[$\mu^+\mu^-$ channel.\label{fig:ams02_a}]
	{\includegraphics[width=.45\textwidth]{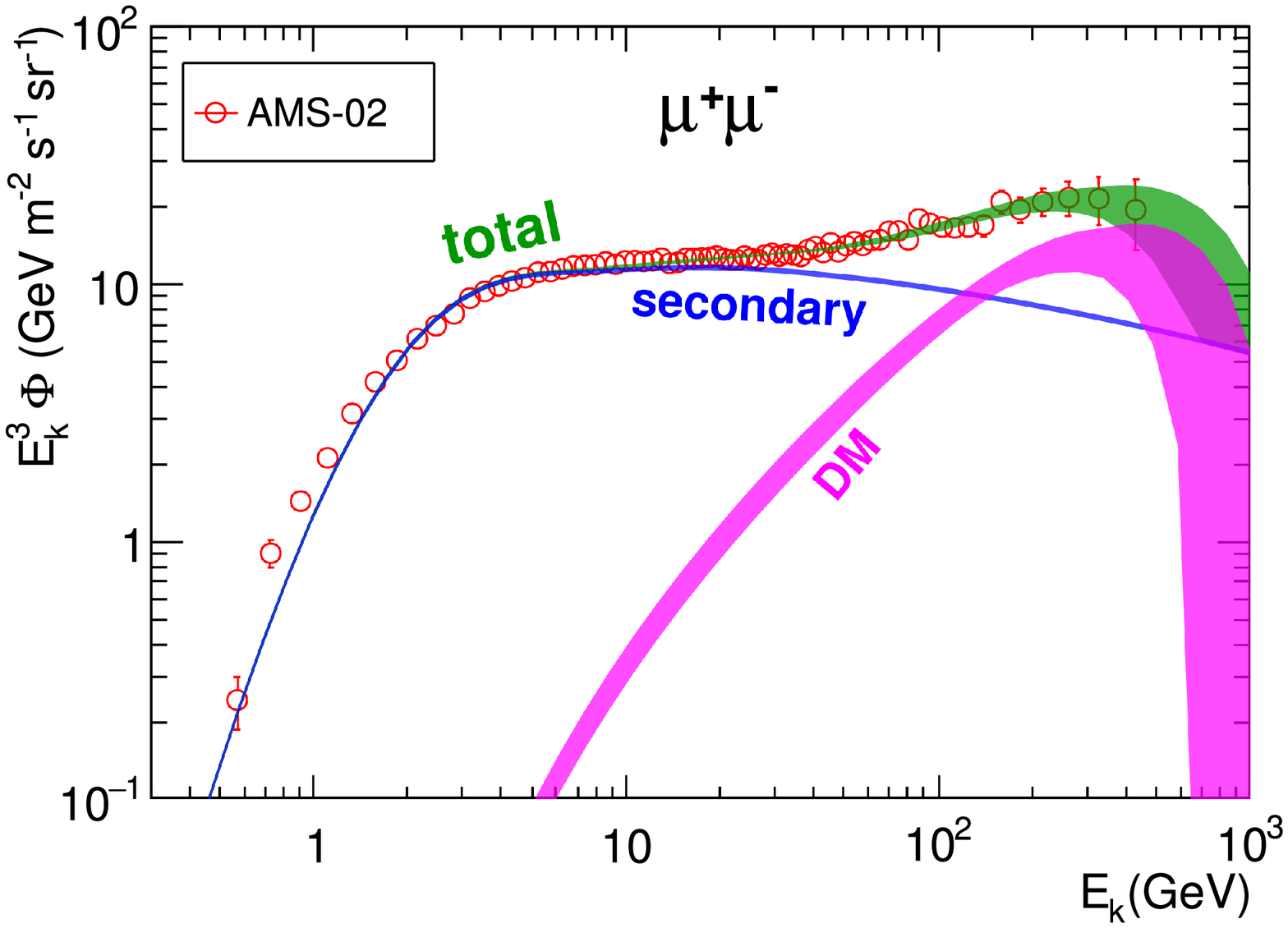}}
	\subfigure[$\tau^+\tau^-$ channel.\label{fig:ams02_b}]
	{\includegraphics[width=.45\textwidth]{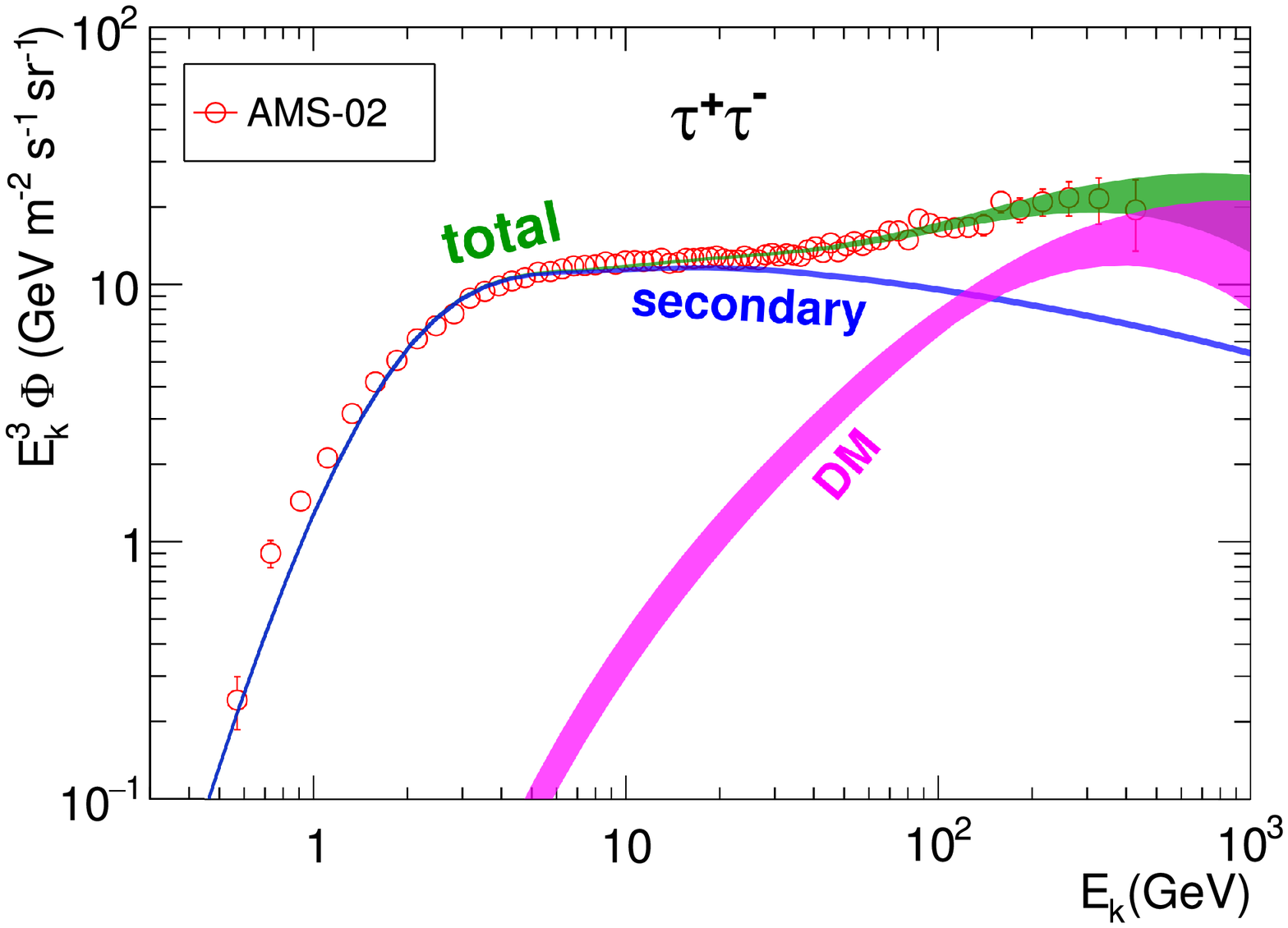}}
  \caption{Fittings to the positron flux measured by AMS-02 for DM annihilations to $\mu^+\mu^-$ (left panel) and $\tau^+\tau^-$ (right panel),respectively. The pink bands indicate the contributions from DM annihilation within $2\sigma$ uncertainty. The blue lines represent the secondary CR positron flux. Total positron fluxes are shown as green bands.
  \label{fig:ams02}
}
\end{figure}

\begin{table}[htb]
\centering
\begin{tabular}{c|c|c|c|c}
\hline\hline
Channels    & $m_\chi (\TeV)$ &  AMS-02 ($2\sigma$)    & Fermi limits    & Planck limits\\
\hline
$\mu^+ \mu^-$ & 0.89 & $3.79 \times 10^{-24} < \sigmav < 6.48 \times 10^{-24}$
               &  $2.95 \times 10^{-24}$
					&  $2.58 \times 10^{-24}$
\\
$\tau^+ \tau-$ & 3.89  & $5.29 \times 10^{-23} < \sigmav < 1.06 \times 10^{-22}$
               &  $ 1.25 \times 10^{-23}$
					&  $ 1.06 \times 10^{-23}$
\\
\hline \hline
\end{tabular}
\caption{The best-fit values of DM masses $m_\chi$ and corresponding thermally averaged annihilation cross sections $\sigmav$ (in units of $\mathrm{cm}^3\mathrm{s}^{-1}$) given by the fitting to the AMS-02 data with the DR propagation model. The corresponding limits from the Fermi-LAT and Planck observations are also shown.
}
\label{tb:limits}
\end{table}

Combining the contributions of primary CR $e^-$, secondary CR $e^\pm$, and $e^\pm$ from DM annihilation, we perform a fit to the latest AMS-02 $e^\pm$ data, including the positron fraction $\frac{e^+}{e^+ + e^-}$ and the fluxes of $e^+$, $e^-$, and $e^+ + e^-$~\cite{AMS02-posi-2014,AMS02-elec-2014,AMS02-tot-2014}.
We provide the fitting results to the observed $e^+$ flux with the DR prorogation model in Fig.~\ref{fig:ams02}; the bands representing $2\sigma$ uncertainties are also shown. The best-fit values of $m_\chi$ and related $2\sigma$ regions of $\sigmav$ (in $\mathrm{cm}^3\mathrm{s}^{-1}$) are listed in Table.~\ref{tb:limits}.
The corresponding exclusion limits derived from the Fermi-LAT dwarf galaxy gamma-ray~\cite{Ackermann:2015zua} and Planck CMB~\cite{Slatyer:2015jla} observations are also given.
It is obvious that the parameter regions of $\sigmav$ for explaining the AMS-02 $e^\pm$ excess are excluded by other two kinds of observations.
Compared to the $\mu^+\mu^-$ channel, the tension in the $\tau^+\tau^-$ channel is severer due to tremendous photons from the hadronic decays of $\tau$.

\section{BREIT-WIGNER ENHANCEMENT}
\label{sec:enhance}

In the Breit-Wigner scenario, the DM annihilation cross section has a typical form of
\begin{equation}
  \begin{split}
  \sigma v \propto \frac{1}{16 \pi m_\chi^2}\frac{1}{(\delta+v^2/4)^2+\gamma^2 }
\end{split}
\label{eq:sigmav_s}
\end{equation}
This form is valid in the non-relativistic limit with $v^2 << 1$ and $\delta<< 1$ at the center-of-mass energy $\sqrt{s} \sim \sqrt{4 m_\chi^2 + m_\chi^2 v^2} $.

As an example, we consider a simple leptophilic fermionic DM model, where DM particles interact with charged leptons through a vector mediator $Z'$ \cite{Bi:2009uj}. The corresponding lagrangian is
\begin{equation}
  \mathscr{L}_{int}  \supset -g(a \bar{\chi} \gamma^\mu \chi +
    \bar{l_i} \gamma^\mu l_i
    ) Z_\mu',
\label{eq:l}
\end{equation}
where $l_i$ represents the species of leptons, $g$ and $a g$ are the couplings of $Z'$ to the leptons and DM particles, respectively.
This model can easily avoid the constraints from DM direct detection and collider experiments due to its leptophilic property.

The DM annihilation cross section in this model is given by
\begin{equation} \label{eq:sv}
  \sigma v = \frac{1}{ 6 \pi}\frac{a^2 g^4 s}{(s-m_{Z'}^2)^2+ m_{Z'}^2 \Gamma_{Z'}^2}
  (1+\frac{ 2m_\chi^2}{s}),
\end{equation}
where $m_\chi$, $m_{Z'}$ and $\Gamma_{Z'}$ are the DM mass, the $Z'$ mass, and the decay width of $Z'$, respectively,
$v$ is the relative velocity between two incident DM particles. Note that the lepton mass has been neglected in Eq.~\ref{eq:sv}
due to the large $\sqrt{s}$ considered in our analysis. The decay width of $Z'$ can be expressed as
\begin{equation}
  \Gamma_{Z'} = \frac{m_{Z'}}{12 \pi} a^2 g^2 \xi_\chi^3 \Theta(m_{Z'}-2 m_{\chi})
  + \frac{m_{Z'}}{12 \pi} g^2  \xi_{l_i}^3,
\end{equation}
where $\xi_\chi \equiv \sqrt{1-4 m_\chi^2 / m_{Z'}^2}$, $\xi_{l_i} \equiv \sqrt{1-4 m_{l_i}^2 / m_{Z'}^2}$, and $\Theta(x)$ is the unit step function.
For $m_{Z'} \sim 2 m_\chi$, $Z'$ dominantly decays to leptons with the decay width given by $\sim g^2 m_{Z'}/{12\pi^2}$.

Then we calculate the thermally averaged DM annihilation cross section through the formula of ~\cite{Gondolo:1990dk}
\begin{equation}
  \sigmav =\frac{1}{n_{EQ}^2} \frac{m_\chi}{ 64 \pi^4 x}
  \int_{4 m_\chi^2}^{\infty} \hat{\sigma}(s) \sqrt{s} K_1(\frac{x\sqrt{s}}{m_\chi}) ds,
  \label{eq:sigmav}
\end{equation}
with
\begin{equation}
  \begin{split}
    n_{EQ} & = \frac{g_i}{2 \pi^2}\frac{m_\chi^3}{x} K_2(x),\\
    \hat{\sigma}(s) & = 2 g_i^2 m_\chi \sqrt{s- 4 m_\chi^2} \sigma v,
  \end{split}
\end{equation}
where $K_i(x)$ is the modified Bessel function of order	$i$, $g_i$ is the internal degree of freedom of the DM particle, which equals 4 in this model.

The evolution of the DM density is determined by numerically solving the Boltzmann equation
\begin{equation}
  \begin{split}
    \frac{d Y}{d x} =  - \frac{ s(x)}{Hx} \sigmav (Y^2 -Y_{eq}^2),
  \end{split}
\end{equation}
 where $Y \equiv n/s$, $n$ is the DM number density, $s = \frac{2 \pi^2 }{45} g_{\ast s}\frac{m^3}{x^3}$ is the Universe entropy density, $H= \sqrt{\frac{4 \pi^3 g_{\ast}}{45 m_{pl}^2}} \frac{m^2}{x^2}$ is the Hubble parameter, and $g_{\ast s}$ and $g_\ast$ are the effective degrees of freedom defined by the entropy density and the radiation density, respectively.

\section{Results}
\label{sec:results}

\begin{figure}[!htbp]
	\subfigure[$\delta < 0$.\label{fig:efactor_a}]
	{\includegraphics[width=.45\textwidth, trim={40 20 40 30},clip]{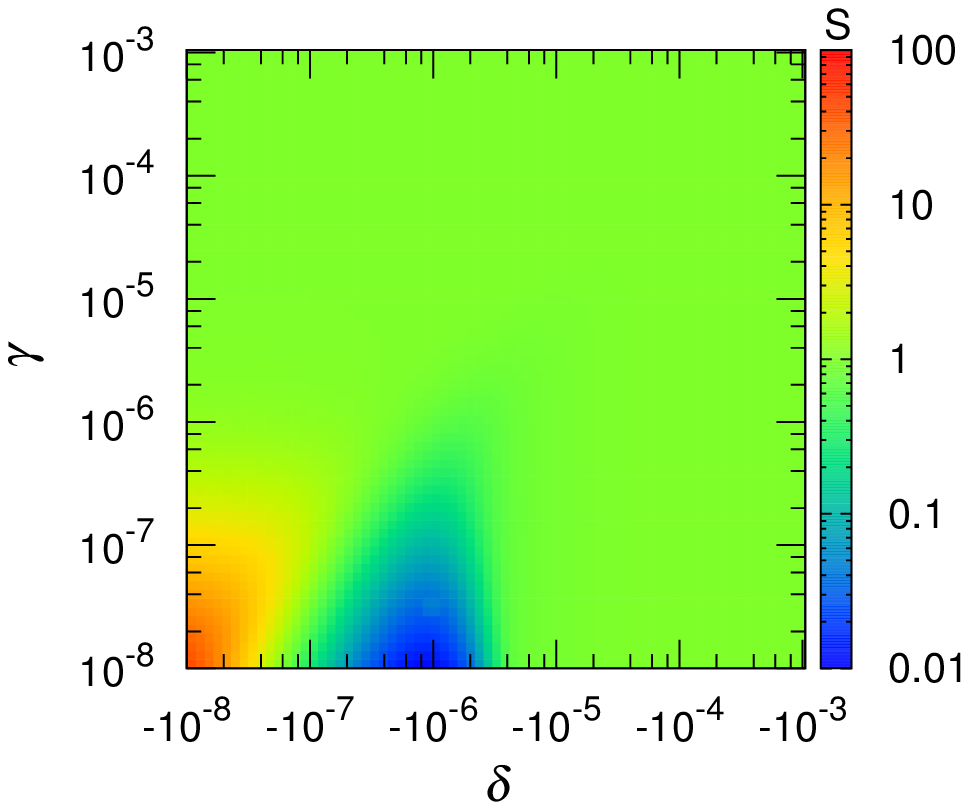}}
	\subfigure[$\delta > 0$.\label{fig:efactor_b}]
	{\includegraphics[width=.45\textwidth, trim={40 20 40 30},clip]{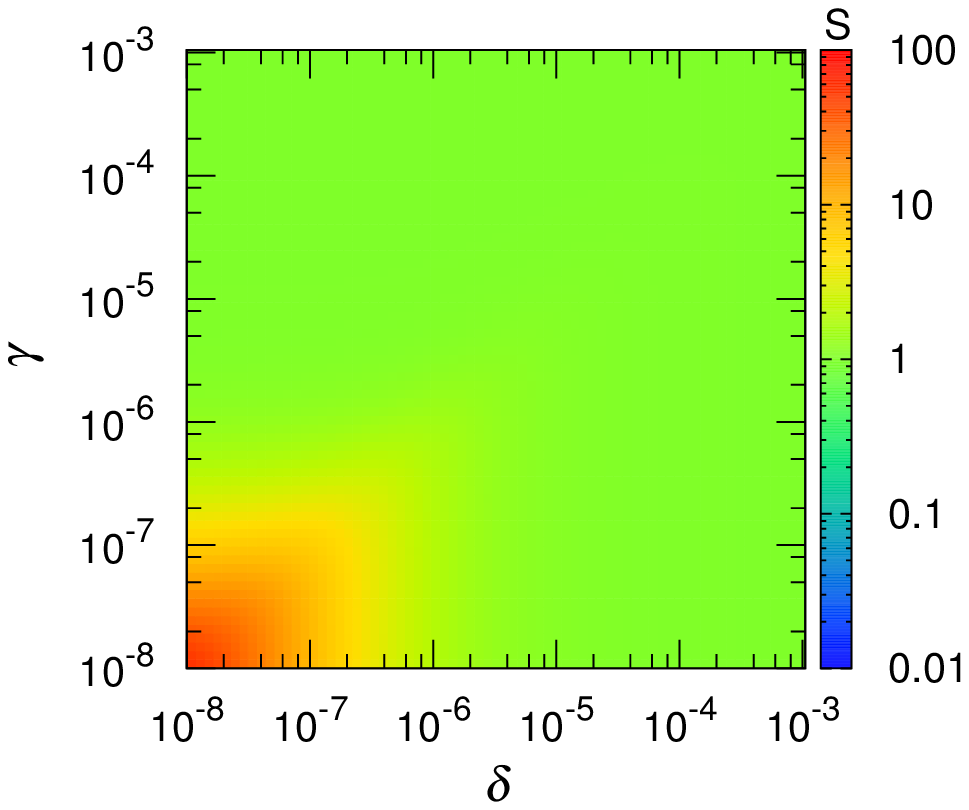}}
  \caption{The scaling factor $S\equiv \sigmav_\mathrm{D}/\sigmav_\mathrm{G}$ in the $\delta-\gamma$ plane, where $\sigma_\mathrm{D}$ and $\sigma_\mathrm{G}$ denote the annihilation cross sections in dwarf galaxies with $v=10^{-4}$ and near the solar system in the Galaxy with $v=10^{-3}$, respectively. The left and right panels represent the physical pole case with $\delta<0$ and unphysical pole case with $\delta>0$, respectively.
  \label{fig:efactor}
}
\end{figure}

In principle, we can accommodate the DM explanations for observations with different DM relative velocities. Only the DM particles located in the Galaxy within a range of $\sim 1$ kpc around the Solar system could provide significant contributions to the observed high energy CR $e^\pm$, because of the prorogation effects. The typical relative velocities of these particles are $\sim 10^{-3}$, while the typical relative velocities of DM particles in dwarf galaxies are $\sim 10^{4}$. Their annihilation cross sections may be very different in the velocity dependent annihilation models. In order to obtain the constraints on $\sigmav_\mathrm{G}$, the constraints on $\sigmav_\mathrm{D}$ from the Fermi-LAT observation should be rescaled by a factor of $1/S \equiv \sigmav_\mathrm{G} /\sigmav_\mathrm{D}$, where $\sigmav_\mathrm{D} $ and $\sigmav_\mathrm{G}$ are the thermally averaged DM annihilation cross sections in dwarf galaxies and near the solar system in the Galaxy, respectively. In order to relax the tension between the DM explanations for the Fermi-LAT and AMS-02 observations, the $S$ factor should be smaller than 1.

We show the $S$ factor in Fig.~\ref{fig:efactor}, and find that a parameter region with $10^{-8} < \gamma < 10^{-6}$ and $-4 \times 10^{-6} < \delta < -10^{-7}$ can satisfy our requirement with $S \ll 1$. For the cases of $\delta > 0$ corresponding to an unphysical pole, there is no parameter region with $S<1$ as shown in Fig.~\ref{fig:efactor_b}. This can be understood by Eq.~\ref{eq:sigmav_s}: the DM annihilation cross section always increases with decreasing relative velocity for $\delta >0$. Therefore, only the cases of $\delta < 0$ can be used to relax the tension between different observations.

\begin{figure}[!h]
	\subfigure[$\mu^+\mu^-$ channel.]
	{\includegraphics[width=.45\textwidth]{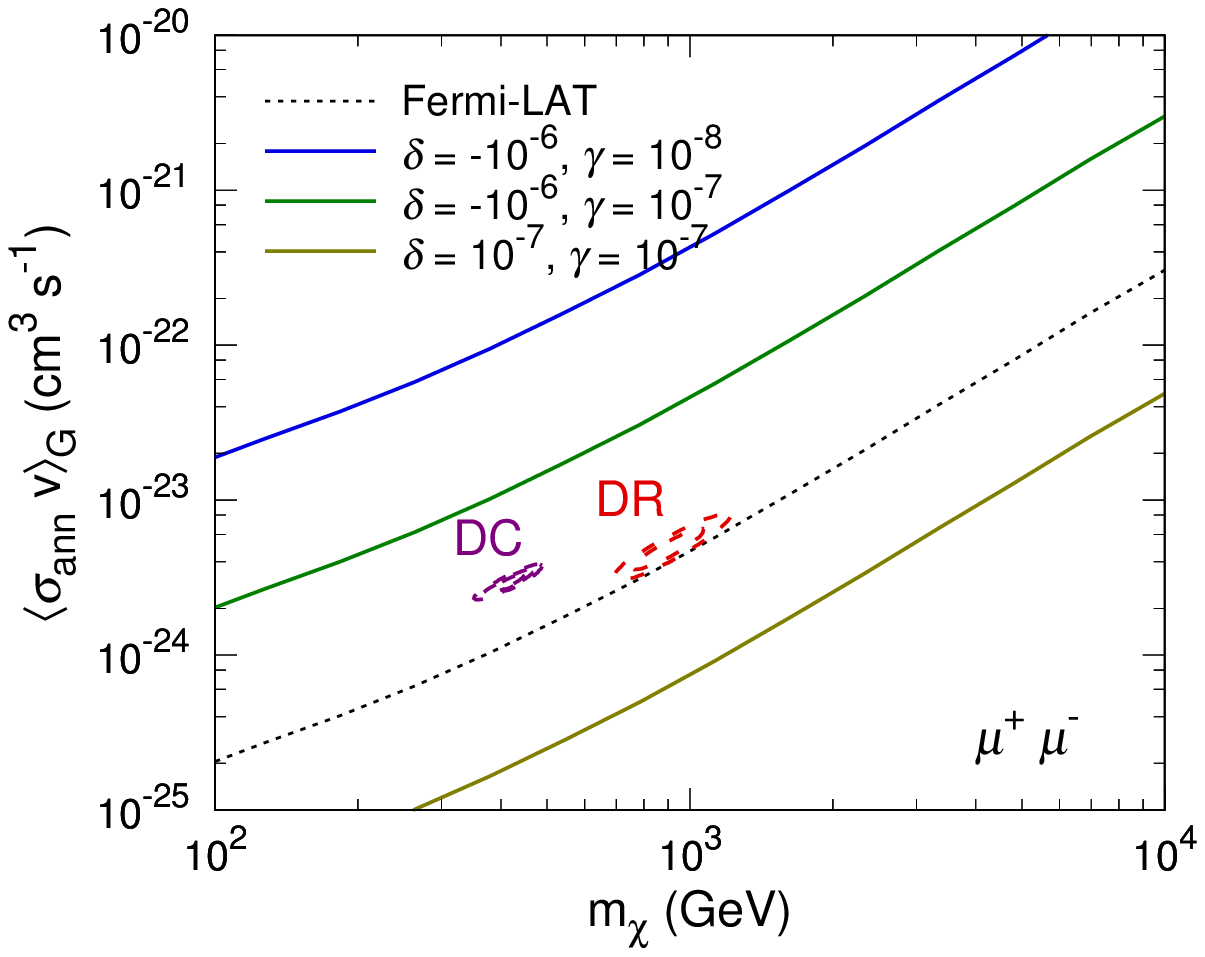}}
	\subfigure[$\tau^+\tau^-$ channel.]
	{\includegraphics[width=.45\textwidth]{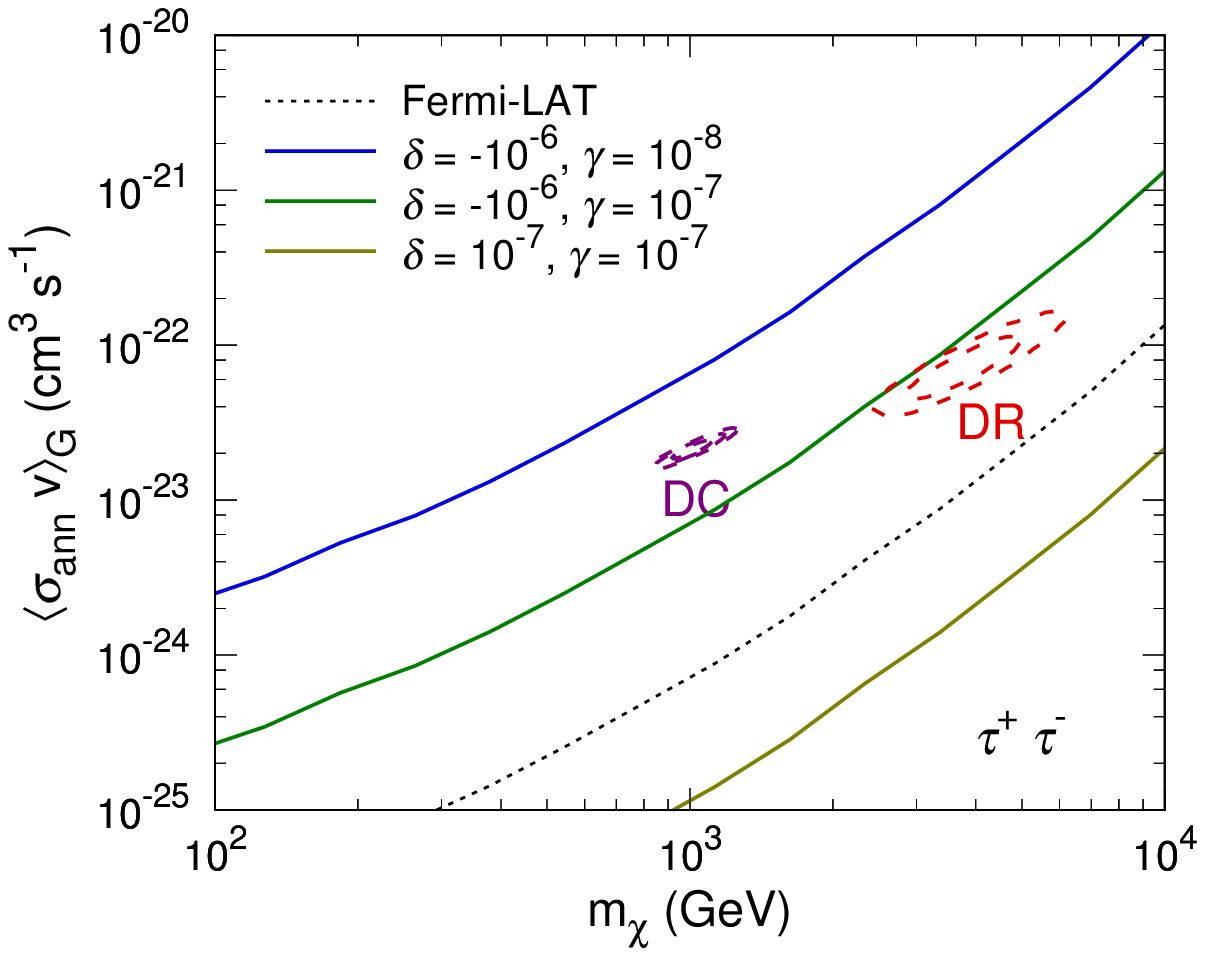}}
  \caption{Contour regions represent the parameter regions accounting for the AMS-02 results in the DC and DR propagation models.
  Solid lines are the constraints on $\sigmav_\mathrm{G}$ from the Fermi-LAT dwarf galaxy gamma-ray observation for different parameter sets of $\delta$ and $\gamma$. The original the Fermi-LAT limits on $\sigmav_\mathrm{D}$ are also shown. The left and right panels represent the cases of DM annihilation to $\mu^+\mu^-$ and $\tau^+\tau^-$, respectively.
  }
\label{fig:fermi}
\end{figure}

In Fig.~\ref{fig:fermi}, we compare the parameter regions accounting for the AMS-02 $e^\pm$ excess with the dwarf galaxy gamma-ray constraints, which are obtained by rescaling the limits given by the Fermi-LAT collaboration \cite{Ackermann:2015zua}. It is shown that the cases with a negative tiny $-\delta \leq 10^{-6}$ can evade the dwarf galaxy constraints. As $\sigmav$ is almost proportional to $1/m_\chi^2$ as can be seen from Eq.~\ref{eq:sigmav}, the ratio of $\sigmav_\mathrm{G} /\sigmav_\mathrm{D}$ is independent of the DM mass. Therefore, the modified dwarf galaxy gamma-ray constraints for different parameter sets of $\delta$ and $\gamma$ are parallel in Fig.~\ref{fig:fermi}. Note that in the above analysis, we fix the DM relative velocity in dwarf galaxies to be $v=10^{-4}$. Strictly speaking, since DM particles in dwarf galaxies have different typical relative velocities with an order of $\mathcal{O}(10^{-4})$, the total constraint should be obtained by combining the individual constraints specified for each dwarf galaxy with a large J factor. A detailed discussion can be found in Ref.~\cite{Zhao:2016xie}.

\begin{figure}[!h]
	\subfigure[$\mu^+\mu^-$ channel.]
	{\includegraphics[width=.45\textwidth]{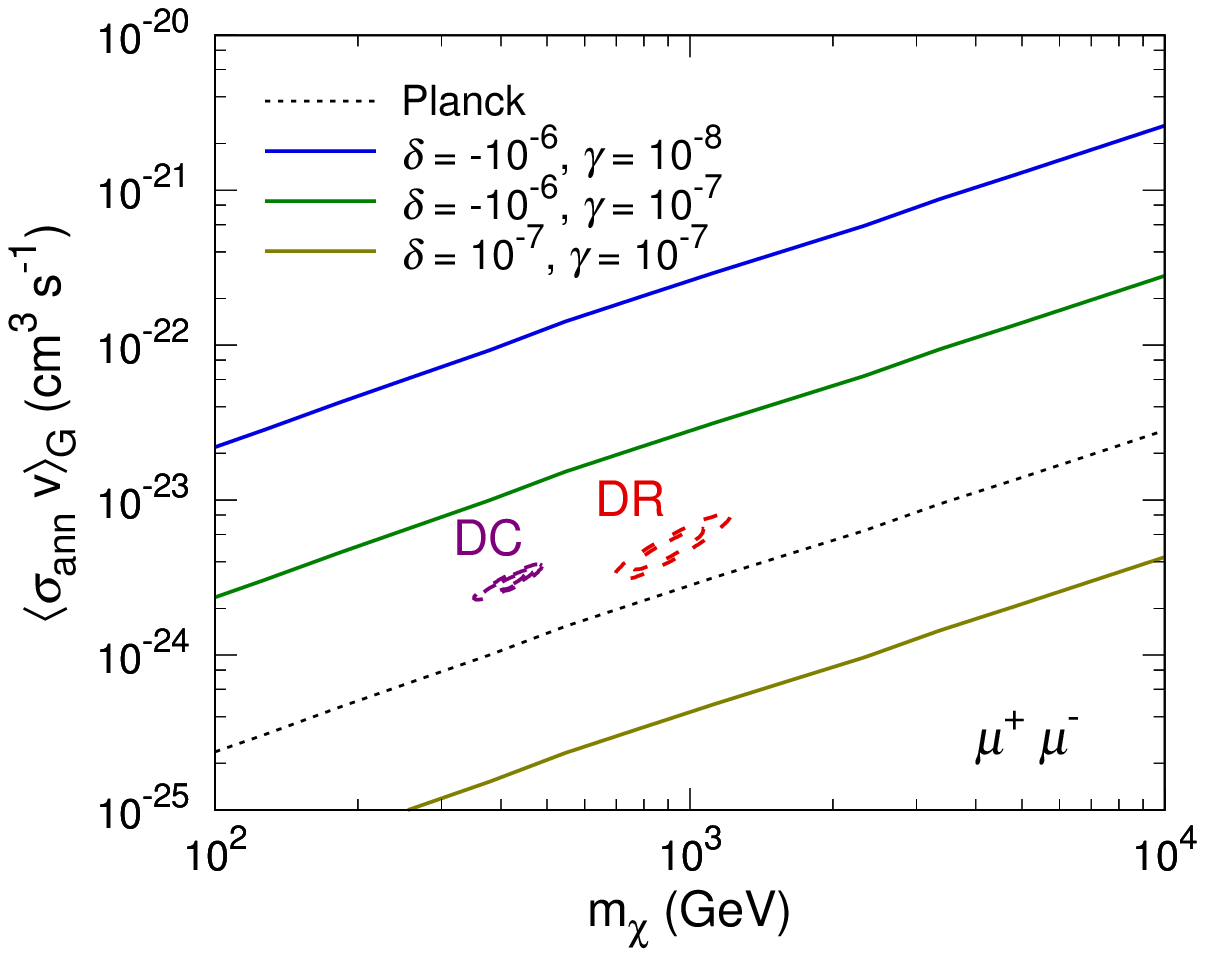}}
	\subfigure[$\tau^+\tau^-$ channel.]
	{\includegraphics[width=.45\textwidth]{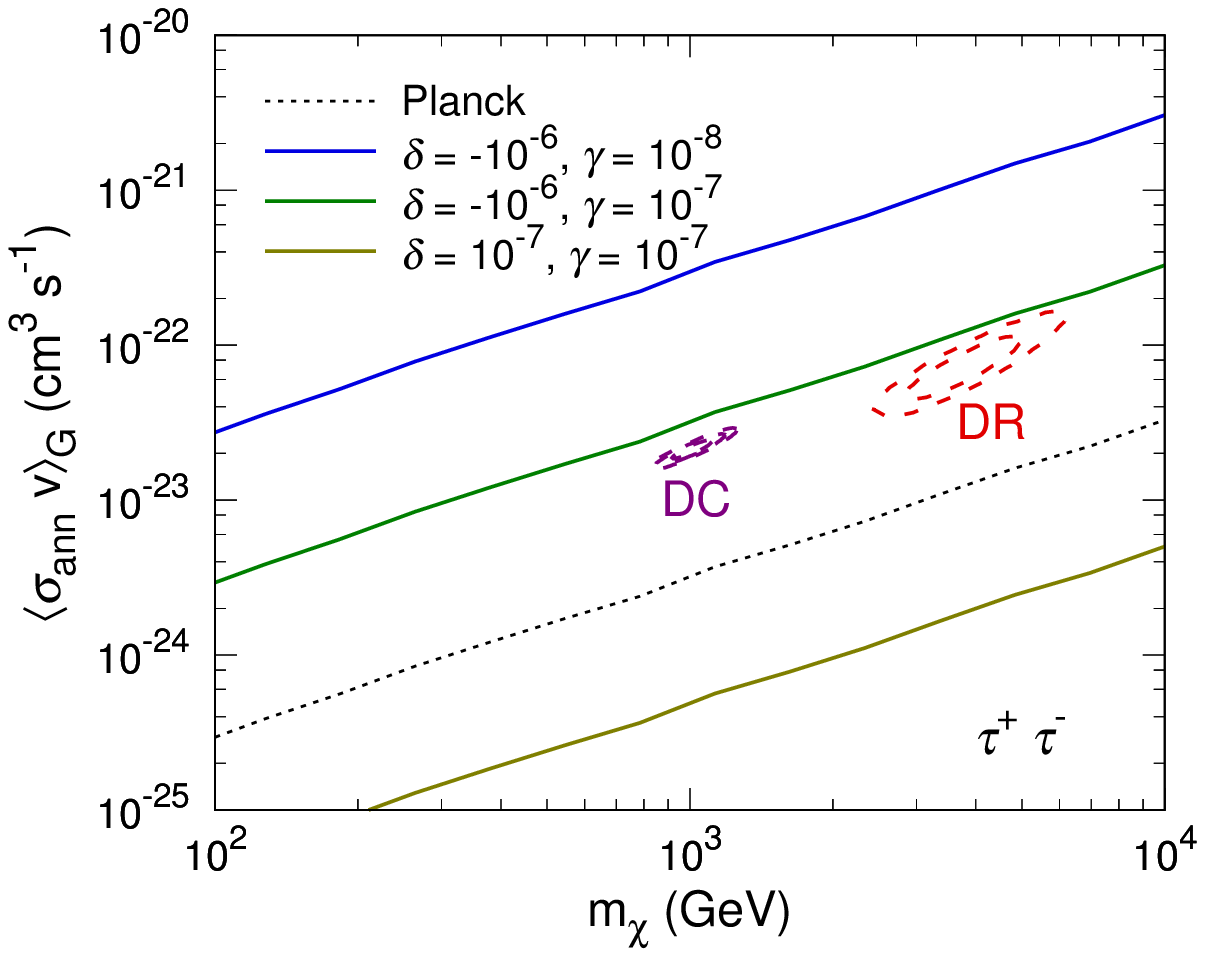}}
	\caption{The same as Fig.~\ref{fig:fermi} but the constraints are derived from the Planck CMB observation.
  }
\label{fig:planck}
\end{figure}

The above analysis can be directly applied to reconcile the tension between the DM explanations for the AMS-02 $e^\pm$ and Planck CMB observations. In order to derive the constraints on $\sigmav_\mathrm{G}$ from CMB observations, we define a rescaling factor of $1/S' \equiv \sigmav_\mathrm{G}/\sigmav_\mathrm{z_r}$, where $\sigmav_\mathrm{z_r}$ is the thermally averaged annihilation cross section of DM particles affecting CMB at recombination with $v\ll 10^{-6}$. In fact, the Breit-Wigner effect would saturate for DM particles with such a small $v$. In Fig.~\ref{fig:fermi}, we compare the parameter regions accounting for the AMS-02 $e^\pm$ excess with the CMB constraints, which are obtained by rescaling the limits given by Ref.~\cite{Slatyer:2015kla}. We find that the cases with a negative tiny $\delta \sim -10^{-6}$ can also evade the CMB constraints.

\begin{figure}[!h]
	\subfigure[$\mu^+\mu^-$ channel.\label{fig:enhance_m_a}]
	{\includegraphics[width=.45\textwidth]{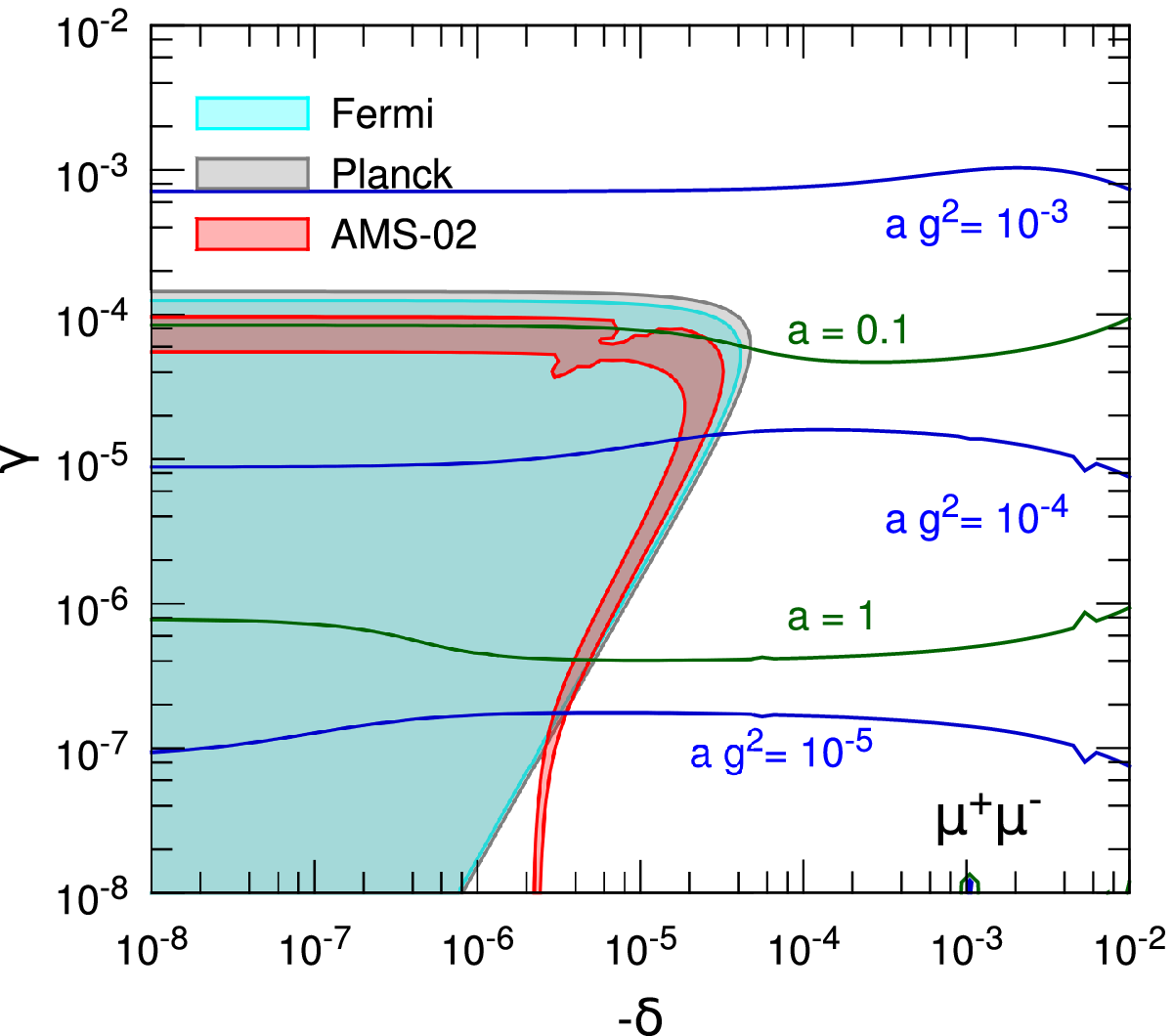}}
	\subfigure[$\tau^+ \tau^-$ channel.\label{fig:enhance_m_b}]
	{\includegraphics[width=.45\textwidth]{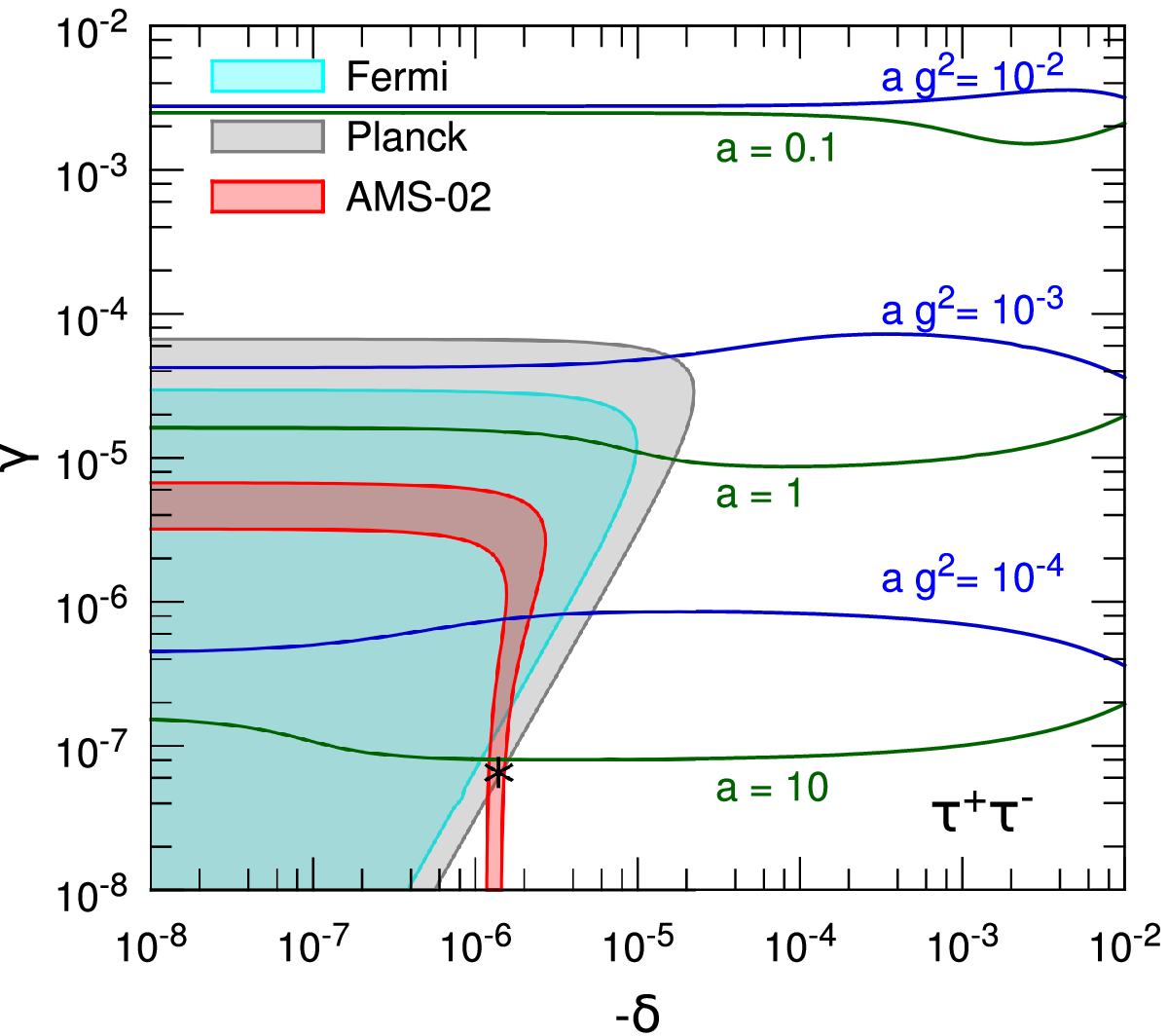}}
  \caption{Parameter regions accounting for various observations in the $\delta-\gamma$ plane with $\delta<0$ for DM annihilation to $\mu^+\mu^-$ (left panel) and $\tau^+\tau^-$ (right panel), respectively. The DM mass is taken to be the value given in  Table.~\ref{tb:limits}. In each parameter point, $a$ and $ag^2$ are derived by requiring the correct relic density $\Omega h^2 =0.1188$; then $\sigmav_\mathrm{G}$, $\sigmav_\mathrm{D}$, and $\sigmav_\mathrm{Z_r}$ can also be obtained. Red shaded region are the parameter regions corresponding to $\sigmav_\mathrm{G}$ given in Tab.~\ref{tb:limits} which can explain the AMS-02 results. The gray and cyan regions denote parameter regions excluded by the Planck and Fermi-LAT observations, respectively. The green and blue solid lines are the isolines of $a$ and $ag^2$, respectively.}
\label{fig:enhance_m}
\end{figure}

For each point in the $\delta-\gamma$ plane with $\delta<0$, we determine $ag^2$ and $a$ through the correct relic density $\Omega h^2=0.1188$ \cite{Ade:2015xua} by resolving the Bolzmann equation, and derive corresponding $\sigmav_\mathrm{G}$, $\sigmav_\mathrm{D}$, and $\sigmav_\mathrm{Z_r}$. In Fig.~\ref{fig:enhance_m}, the red bands represent the parameter regions simultaneously accounting for the AMS-02 CR $e^\pm$ excess and the correct relic density. Here we only consider $m_\chi$ and $\sigmav_\mathrm{G}$ derived with the DR propagation model as given in Table.~\ref{tb:limits}. The parameter regions excluded by the Fermi-LAT and Planck limits are also shown in Fig.~\ref{fig:enhance_m}. We find that there exists a parameter region with $\gamma \uwave{<}  10^{-7}$ and $\delta \sim -10^{-6}$, which can accommodate all the observations.

We also show the isolines of $a g^2$ and $a$ satisfying the correct DM relic density in Fig.~\ref{fig:enhance_m}. The behavior of these lines can be understood as follows. Roughly speaking, the thermal relic density $\Omega h^2$ in the usual DM models is determined by the freeze-out temperature $x_f \sim \mathcal{O}(10)$ (corresponding to $v^2 \sim 10^{-1}$) and $\sigmav_\mathrm{f}$ as $\Omega h^2 \propto x_f /\sigmav_\mathrm{f}$. For the resonant case, since the annihilation cross section would increase with dropping temperature, the annihilation process may be significant until the Breit-Wigner effect almost saturates at a temperature of $x_b$. $x_b$ can be roughly determined by $|\delta|^{-1}$ for $\delta< 0$. This is because that there are not enough DM particles with velocities of $v\sim |\delta|^{\frac{1}{2}}$ for sufficient resonant annihilation when $x\gg 1/|\delta|$. Using the approximated form of $\sigmav_\mathrm{b} \propto a^2g^4 |\delta|^{\frac{1}{2}} x_b^{\frac{3}{2}} /\gamma$ by integrating out the pole~\cite{Bai:2017fav}, we get $\Omega h^2 \propto x_b /\sigmav_\mathrm{b} \propto \gamma/a^2g^4$. Therefore, the correct relic density can be easily obtained by adjusting $ag^2$ with $\gamma^{\frac{1}{2}}$ as shown in Fig.~\ref{fig:enhance_m}.

An important issue that should be addressed is the kinetic decoupling effect. In the parameter regions discussed above, since the scatterings between DM particles and SM radiations are not sufficient due to the t-channel exchange of a heavy mediator, the kinetic decoupling would occur at a high temperature of $T > \mathcal{O} (1)$ GeV. The velocities of DM particles drop as $\sim R^{-1}$ after the kinetic decoupling rather than $\sim R^{-\frac{1}{2}}$ before the kinetic decoupling, where $R$ is the scale factor of the Universe. Then the Breit-Wigner mechanism would significantly enhance the DM annihilation cross section at the freeze-out epoch and drastically reduce the DM relic density. As discussed in Ref.~\cite{Bi:2011qm}, it is difficult to simultaneously explain the CR $e^\pm$ excess and the relic density with such a significant kinetic decoupling effect. Moreover, after the kinetic decoupling, the velocity distribution of DM particles would depart from the thermal distribution and is difficult to deal with in the calculation of the relic density. A solution is introducing some additional mediators, which can enhance the scattering rate between DM particles and SM radiations and/or the DM self scattering rate. The detailed discussions can be found in Ref.~\cite{Feng:2010zp,Bai:2017fav}.

\section{CONCLUSIONS and DISCUSSIONS}
\label{sec:conclusion}

In this work we show that the DM annihilation througth the Breit-Wigner mechanism can reconcile the tension between the DM explanation for the AMS-02 CR $e^\pm$ excess and the constraints from Fermi-LAT dwarf galaxy gamma-ray and Planck CMB observations. Since DM particles affecting these observations have different relative velocities, their annihilation cross sections are different for interpretating the experimental results. In order to check whether the DM explanation for the AMS02 results is excluded by other observations, we should translate all the limits into those on $\sigmav_\mathrm{G}$ for DM particles with a typical relative velocity $v\sim 10^{-3}$.

We take a leptophilic $Z'$ model as a benchmark model. This kind of leptophilic model is not constrained by the results of current direct detection and collider experiments. For the tiny values of the mediator decay width and the mass deviation from the pole, $\sigmav$ would be sensitive to the relative velocity. For the unphysical pole case with $\delta > 0$, $\sigmav$ increases with deceasing velocity. Thus the enhanced constraints on $\sigmav_\mathrm{G}$ from the dwarf galaxy gamma-ray and CMB observations exclude the explanation for the CR $e^\pm$ excess in this case.

For the physical pole case with $\delta \sim -10^{-6}$, DM particles accounting for the CR $e^\pm$ excess with $v\sim \mathcal{O}(10^{-3})$ have the largest annihilation cross section close to the pole. On the other hand, the DM annihilation cross section is suppressed for DM particles with smaller relative velocities in dwarf galaxies and at recombination, which may impact on the gamma-ray and CMB observations, respectively. Therefore, the constraints on $\sigmav_\mathrm{G}$ from these observations are weaken. We find that a parameter region with $\delta \sim -10^{-6}$ and $\gamma \lesssim 10^{-7}$ can simultaneously account for the AMS-02, Fermi-LAT dwarf galaxy gamma-ray, and Planck CMB observations, and the relic density.

From the perspective of model building, a question is how to naturally realize the tiny values of $\delta$ and $\gamma$ derived in above analysis. Here we consider the benchmark point with $\gamma=7.1 \times 10^{-8}$ and $\delta= - 1.5 \times 10^{-6}$ marked in the right panel of Fig.~\ref{fig:enhance_m}. For the small decay width of the mediator, we get $g\sim 1.8\times 10^{-3}$ and $ag\sim 1.8 \times 10^{-2}$. These values are easy to realize in a realistic model. The problem is how to achieve a tiny $\delta \sim -10^{-6}$, which seems to require a significant fine-tuning. A solution is given by Ref.~\cite{Bai:2017fav} through the nontrivial flavour symmetry-breaking in the dark sector. By assigning a particular symmetry-breaking mode, a resonance with a mass of almost $2 m_\chi$ can be realized. The tiny mass deviation of $\delta$ is naturally induced by loop effects.

\section*{Acknowledgment}
This work is supported by the National Key Program for Research and Development (No.
2016YFA0400200) and by the National Natural Science Foundation of China
under Grants No. 11475189 and 11475191.

\appendix

\bibliographystyle{JHEP}
\bibliography{RELIC}
\end{document}